\documentclass[twocolumn]{aastex62}

\usepackage{graphics,epsf}
\usepackage{amsmath}                
\usepackage{amsfonts}               
\usepackage{amssymb}                
\usepackage{epsfig}
 \usepackage{epstopdf}
\usepackage{multirow}
\usepackage[para,online,flushleft]{threeparttable}

\def \msyr{~\rm{M_{\odot}}~\rm{yr^{-1}}}

\def \K{~\rm{K}}

\def \kpc{~\rm{kpc}}
\def \etc{$\eta$~Car}

\def \rmModot{~\rm{M_{\sun}}}

\def \rmLodot{~\rm{L_{\sun}}}

\begin{document}

\title{Periodicity in the Light Curve of P~Cygni -- Indication for a Binary Companion?}

\author[0000-0002-1361-9115]{Amir M. Michaelis}
\affil{Department of Physics, Ariel University, Ariel, POB 3, 40700, Israel}
\email{amir.michaeli@msmail.ariel.ac.il}
\author[0000-0002-7840-0181]{Amit Kashi}
\affil{Department of Physics, Ariel University, Ariel, POB 3, 40700, Israel}
\email{kashi@ariel.ac.il}
\author[0000-0001-5249-4354]{Nino Kochiashvili}
\affil{Ilia State University, E. Kharadze Abastumani Astrophysical Observatory, Kakutsa Cholokashvili ave. 3/5, Tbilisi, 0162, Georgia}
\email{nino.kochiashvili@iliauni.edu.ge}

\shorttitle{Periodicity of P~Cyg -- Indication for a Binary?} 
\shortauthors{A.M. Michaelis et al.}

\begin{abstract}
We use observations of the Luminous Blue Variable (LBV) P~Cygni spanning seven decades, along with signal processing methods, to identify a periodicity in the stellar luminosity. 
We find a distinct period of $4.7 \pm 0.3$ years together with shorter periods.
The periodicity is a possible indication of a binary companion passing in an eclipse-like event from the dense LBV wind, and if so it is the first observational indication that P~Cygni is a binary system. This may support models that contribute giant LBV eruptions to interaction with a binary companion.
We discuss other interpretations for the periodicity as well.
\end{abstract}


\section{INTRODUCTION}
\label{sec:intro}

In the year 1600 the Luminous Blue Variable (LBV) P~Cygni (P~Cyg) experienced a major eruption \citep{deGroot1969,deGroot1988}, also known as a ``Supernova Impostor'' (e.g., \citealt{DavidsonHumphreys2012}).
Recalling that the historical event precedes the invention of the telescope,
the eruption (at the time referred to as a nova) caused the star to increase in visible magnitude from below detection by naked eye to $\simeq 3$.
Later on in the seventeenth century the star underwent a series of 4 more eruptions, with decreasing time intervals between them.

P~Cyg was traditionally considered to be a single star.
Even though P~Cyg is the closest LBV to us (at a distance of $1.7 \pm 0.1 \kpc$ ; \citealt{Najarroetal1997}), no companion has ever been observed.
After realizing the progenitor is an LBV, its eruptions were associated with a single star processes (e.g., \citealt{HumphreysDavidson1994,LamersdeGroot1992}).

These kinds of pre-supernova eruptions are thought to occur in the final evolutionary stages of a star.
The best investigated example of a very massive star that had gone through such eruptions and survived is \etc \citep{HumphreysMartin2012}.
But the latter is at least $90 \rmModot$, and probably twice this value \citep{KashiSoker2016Massive}, while P~Cyg is only $\approx 25 \rmModot$
(though we note that there are higher estimates for the mass, such as the one of 
\cite{ElEidHartmann1993} who suggested that stellar evolution tracks support a $50 \rmModot$ star, and \citealt{Lamersetal1983a, Lamersetal1983b}
who favored a $60$--$80 \rmModot$ star).

The peculiar morphology of the nebula which was formed by the eruption of P~Cyg
\citep{Notaetal1995} lead \cite{IsraeliandeGroot1999} to suggest that a different physical process is responsible to the eruptions of $\eta$~Car and P~Cyg,
though the details of such a process were not investigated.

\cite{Kashietal2010} showed that the eruption of P~Cyg lies on a
strip in the total energy vs. timescale diagram (ETD) together with other intermediate
luminosity optical transients, including the two nineteenth century eruptions of \etc.
\cite{KashiSoker2010b} suggested that the same physical mechanism that is applicable to the giant eruptions of LBV stars applies to the other transients in the ETD, including P~Cyg: accretion onto a main-sequence (MS) companion star and release of gravitational energy.

\cite{Kashi2010} explained the series of eruptions of P~Cyg by mass transfer to a B-type binary companion in an eccentric orbit.
He assumed that the luminosity peaks occurred close to periastron passages, as
at these times mass was accreted by the companion and liberated gravitational
energy, part of which went to an increase in luminosity.
\cite{Kashi2010} suggested that mass transfer of $\approx 0.1 \rmModot$ to a B-type binary
companion of $\approx 3$--$6 \rmModot$ can account for the energy of the
eruption, and for the continuously decreasing time interval between the peaks in
the visual light curve of the eruption.
Such a companion was predicted to have an orbital period of $\approx 7$~years, and it was calculated that its Doppler shifts should be detectable with high resolution spectroscopic observations.

An early attempt to find a periodicity in the observations of P~Cyg has been performed by \cite{Israelianetal1996}, who suggested a period of $206 \pm 11$ days.
This periodicity was found in spectra of \ion{Fe} {3} lines.
\cite{Richardsonetal2011} performed a spectroscopic analysis over a period of 15 years but found no periodic radial velocity variation. As they state, the radial velocity variation in the H$\alpha$ line they observed cannot be caused by the companion as the line is formed in a volume much larger than the semi-major axis of the companion predicted by \citet{Kashi2010}.
\cite{Richardsonetal2013} count the non-detection of \cite{Richardsonetal2011} as an argument disfavoring the existence of the companion, but this is inconsistent with the statement of \cite{Richardsonetal2011} regarding the large H$\alpha$ volume.

Recently, \cite{Kochiashvilietal2018} used unpublished observations of P~Cyg obtained by Kharadze and Magalashvili at the Abastumani Observatory to deduce a number of quasi-periods:
($1480 \pm 31$) days; ($736 \pm 27$) days; ($1123 \pm 36$) days; $\sim 579$ days and  $\sim128.7$ days.
The reason for the quasi-periodicity was not discussed in their paper.

In this paper we use photometric observations from the last $2/3$ century in an attempt to recover a periodic signal.
Our premise is that if such a signal exists it would be buried in the data, and if a companion star is present in an orbit of a few years, it is likely
obscured by the high density LBV wind for most of the orbit, and visible only for a short time.

\section{Observations and Analysis}
\label{sec:observations}

We use photometric observations taken from the American Association of Variable Star Observers (AAVSO, \citealt{Kafka2018}).
In addition we use two sets of observations described in \cite{Kochiashvilietal2018}.
Those observations were obtained by Nino Magalashvili and Eugene Kharadze using the 33 cm and 48 cm reflectors of the Abastumani Astrophysical observatory during 1951--1983. They used 29 Cyg and 36 Cyg as comparison and check stars, and obtained two sets of observations corresponding to these two references.

The accuracy of the AAVSO data is $\approx 0.01$ mag in the V band.
This translates to a precision of about $1\%$ in the flux measurements.
However, at most of the nights there are multiple observations, taken by different observers. The average of these observations increases the precision to $0.1$--$0.5 \%$.
Assuming P~Cyg is binary system with two stars with the masses quoted above,
we take for the LBV $T_1 = 18\,200 \K$ and $L_1=5.6 \times 10^5 \rmLodot$ \citep{Najarroetal1997}, and for the companion MS star we take the most favorable values from \citep{Kashi2010} to allow detection $T_2 = 19\,000 \K$ and $L_2= 1\,500 \rmLodot$.
Calculating black-body emission, the expected ratio in the intensity in the visible is $\approx 0.3 \%$.
We thus get an estimation for the magnitude of variation which may be detected using V-Band filter observation.
We therefore conclude that only for optimistic parameters the AAVSO data is of about the required precision to be used in our analysis.
We nevertheless proceed with the analysis with the hope of detecting a binary signal.
The other observations we use are of much higher quality and can therefore be used with no concern.

We first have to join the P~Cyg photometric data from the three sources into one coherent dataset.
To do so, we average same-night observations to obtain a single observation per-night, for each source.
We zero pad the signal at times were no observations have been taken.
Namely, for nights with no data we take $\Delta V =0$.
The next step is to apply a median filter to each signal (per source).
We then re-normalize the data using the following technique. We identify similar measurement points for the three sources, and use them to normalize all data.
We do that by re-quantifying the data to generate a normalized unified dataset that has one point for each night.
All our following analysis is done on this unified, renormalized dataset from our three sources. We hereafter refer to it as the unified signal.

\begin{figure*}
\includegraphics[width=0.95\textwidth]{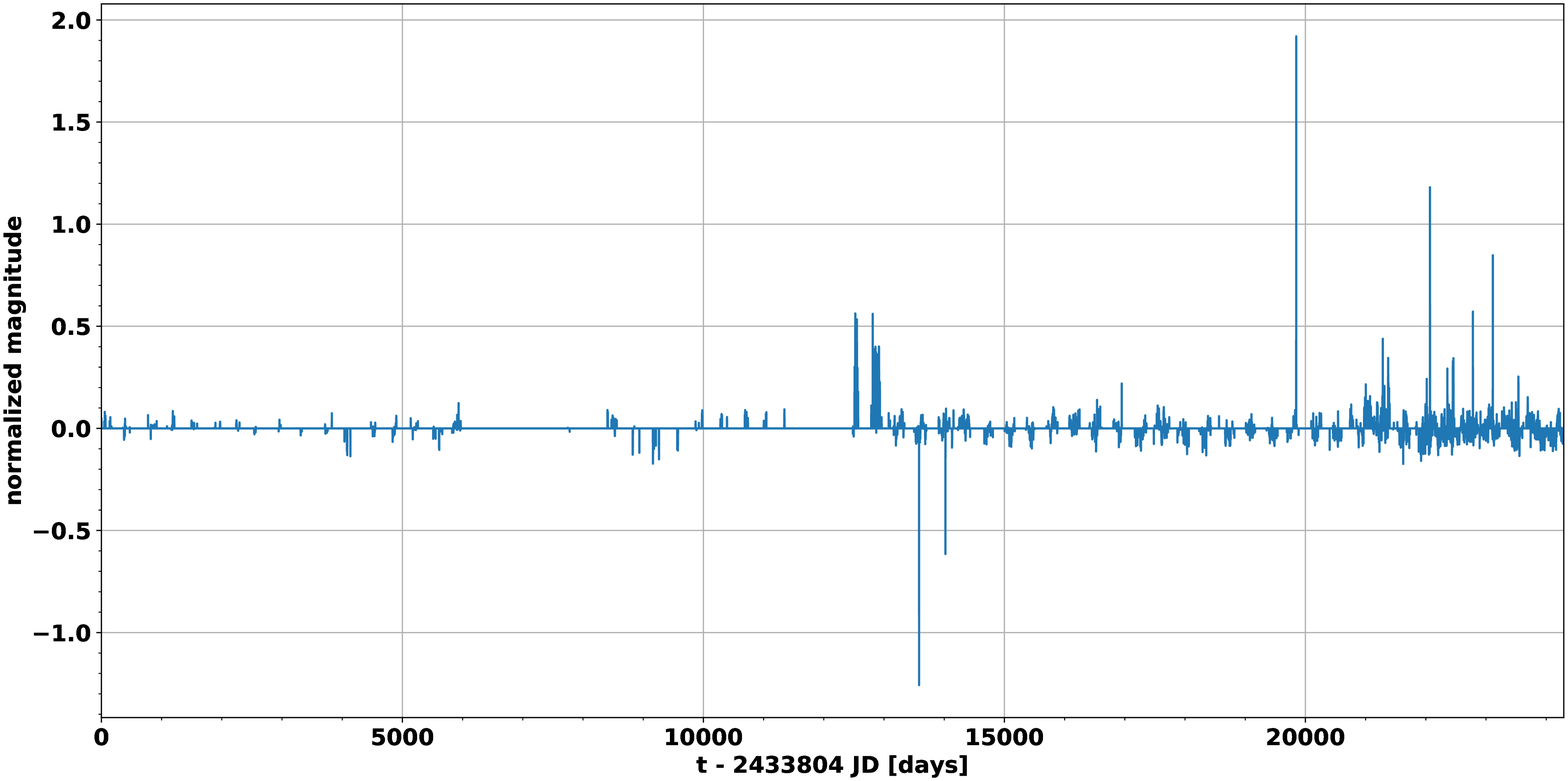}
\includegraphics[width=0.95\textwidth]{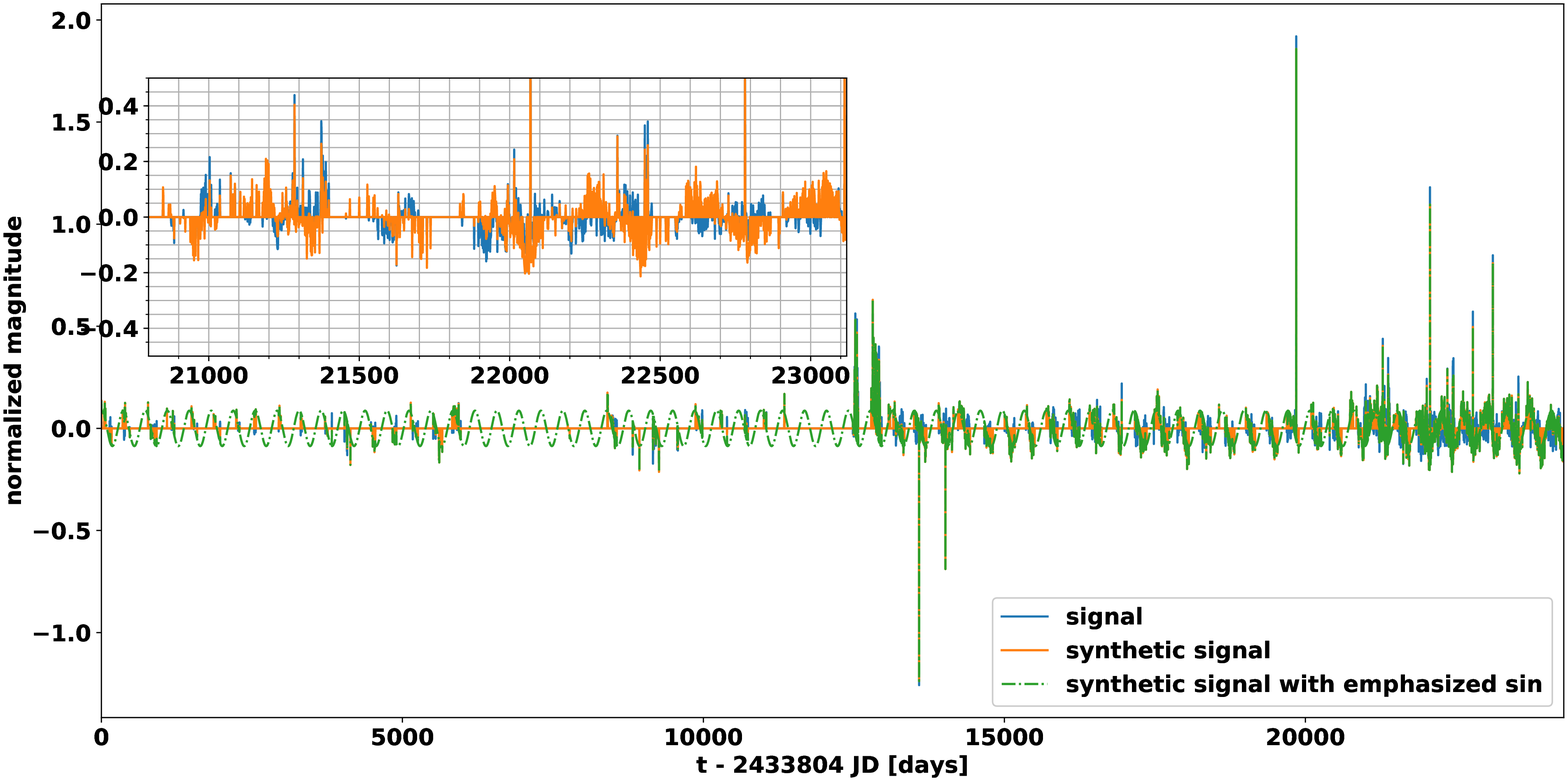}
\caption{Upper panel: normalized magnitude per day from all three data sets. We use data set normalized to 29~Cyg and 36~Cyg (July 1951 -- September 1983) and AAVSO (August 1972 -- December 2017). We normalize the data using the overlap time range, merge it to obtain unified data and create a data set with continuous sample rate of 1 day.\\
Lower panel: We add a synthetic signal that simulates measurements similar to the original signal, with 1~year period. In the figure we add the full values of the 1~year sinusoidal signal for clarity. This signal is used to validate our analysis.
}
\label{fig:m_vs_t}
\end{figure*}

Next, we analyze the frequency spectrum using two methods:
\begin{enumerate}
\item Performing a conventional Fourier transformation using the Fast Fourier Transform (FFT) algorithm.
\item Calculating the power spectrum density (PSD), defined as the spectral power of the auto-correlation of the unified signal.
\end{enumerate}
In order to validate our methods, we add a synthetic signal with a period of 1 year and intensity equal to the variance of the unified signal, and recover it using each of the two methods.
In the upper panel of Fig. \ref{fig:m_vs_t} we show the unified signal.
The time axis is in days starting June 6 1951 (JD~2433804) and contains about 66 years (24292 days) of measurements.
The vertical axis is V-magnitude relative to the data as described above (unified signal).
The lower panel shows the synthetic signal, defined as the unified signal with the added 1~year period signal.
At times where no data is available for the unified signal we did not add the 1~year period signal to our analysis.
The inset zooms on part of the signal to illustrate the way the synthetic signal was constructed.
From the knowledge of the synthetic period we can reverse the analysis process to get a perspective of what we look for in our analysis and how it should be seen.
We use it to ensure the correctness of the analysis and as a prove our work methods.
%

\begin{figure*}
\includegraphics[width=1\textwidth]{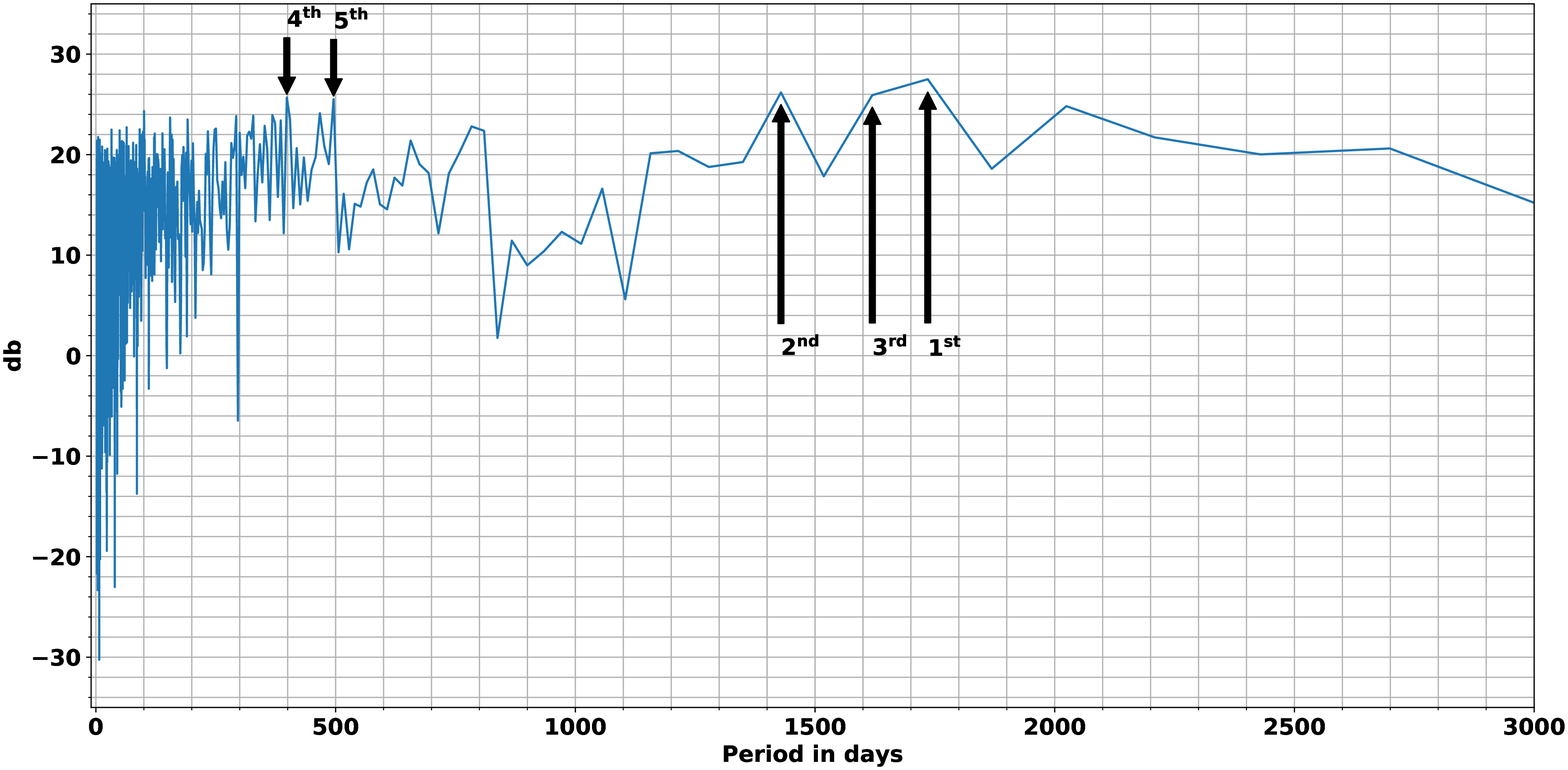}
\includegraphics[width=1\textwidth]{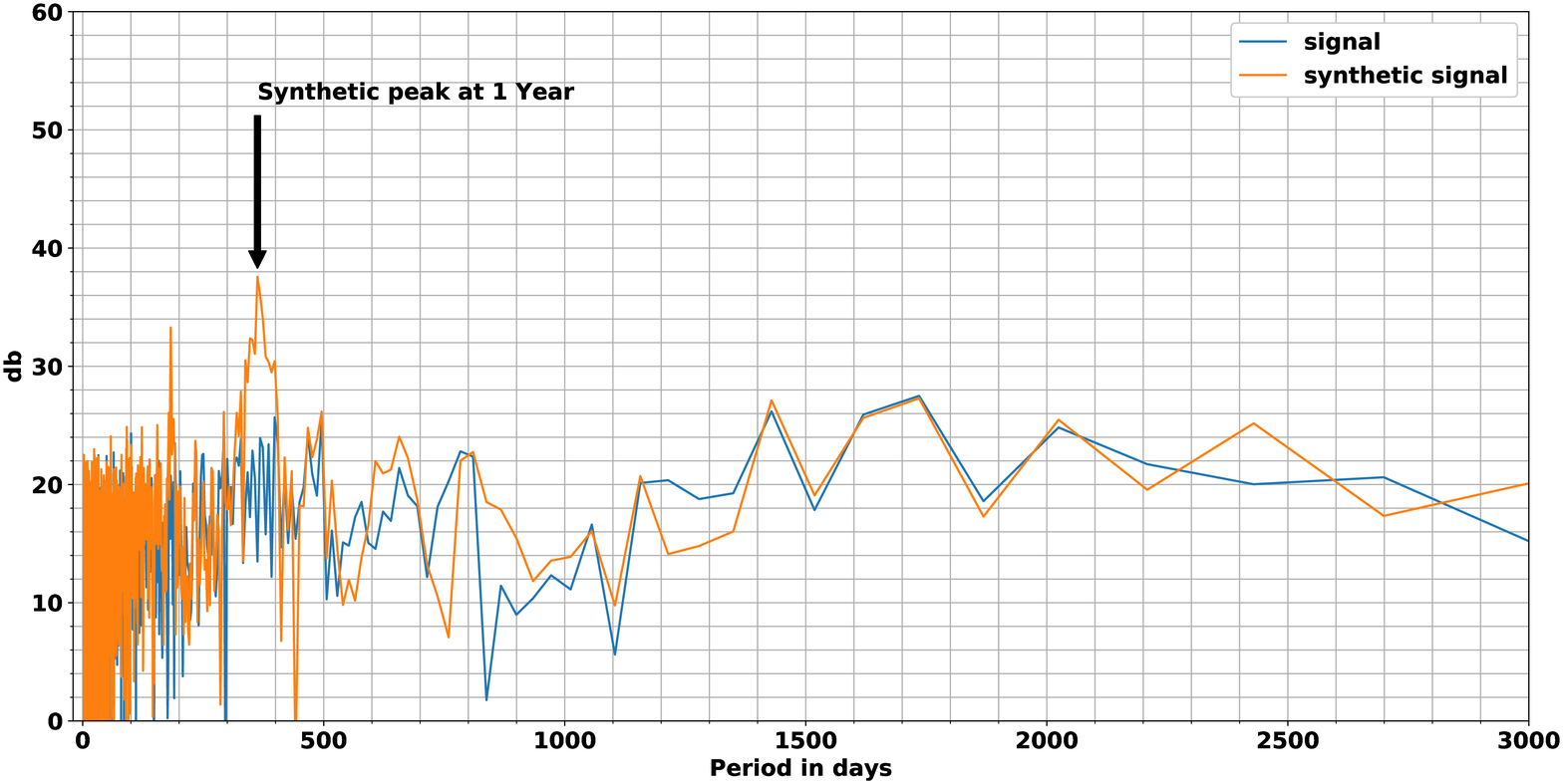}
\caption{Upper panel: spectrum where the x axis is in days and the y axis in db (i.e. $20\log_{10}\left( \left| m(\nu) \right| \right)$.  We regularized our data to a sample rate of 1/day. As a result we can see periods starting from 2 days up to $\simeq 33$ years. As expected we see a lot of variability in the high frequency region (low period). From $P_1=1735$ days onward to short frequencies we see a continuous decline in the signal.\\
Lower panel: We see the influence of a 365 days periodic signal on the analysis. The strong peak represents a simulation of synthetic sinusoidal signal is $\sim37$ db while the maximum value of the signal is seen at $P_1=1735$ days with about $27.5$ db followed by peaks weaker by $1$--$2$~db.
}
\label{fig:spec}
\end{figure*}

In Fig. \ref{fig:spec} we see the spectrum of the unified signal obtained from our FFT analysis.
We find different periods in the signal: $P_1=1735 \pm 115$ days, $P_2=1428 \pm 79$ days, $P_3=1619 \pm 101$ days, $P_4=398 \pm 7$ days and more, in a descending order of strength.
We notice that although we have higher resolution in the low frequencies we find there distinct peaks which differ from their surroundings.
We can clearly see that the synthetic signal produced both the peaks of the unified signal ($27.5$ db at $1735$ days), and the 1~year period signal with power of $37.5$ db. The signal is very strong.
We calculated the statistical properties of the FFT intensity of the synthetic signal and found the mean intensity (the absolute value of the real and complex parts of the FFT) to be 4.2 and the standard deviation to be 2.6.
In this linear scale the intensity of the 1~year peak is 75.6.
This very distinct synthetic signal gives perspective to the other periodicities we find.

%

\begin{figure*}
\includegraphics[width=1\textwidth]{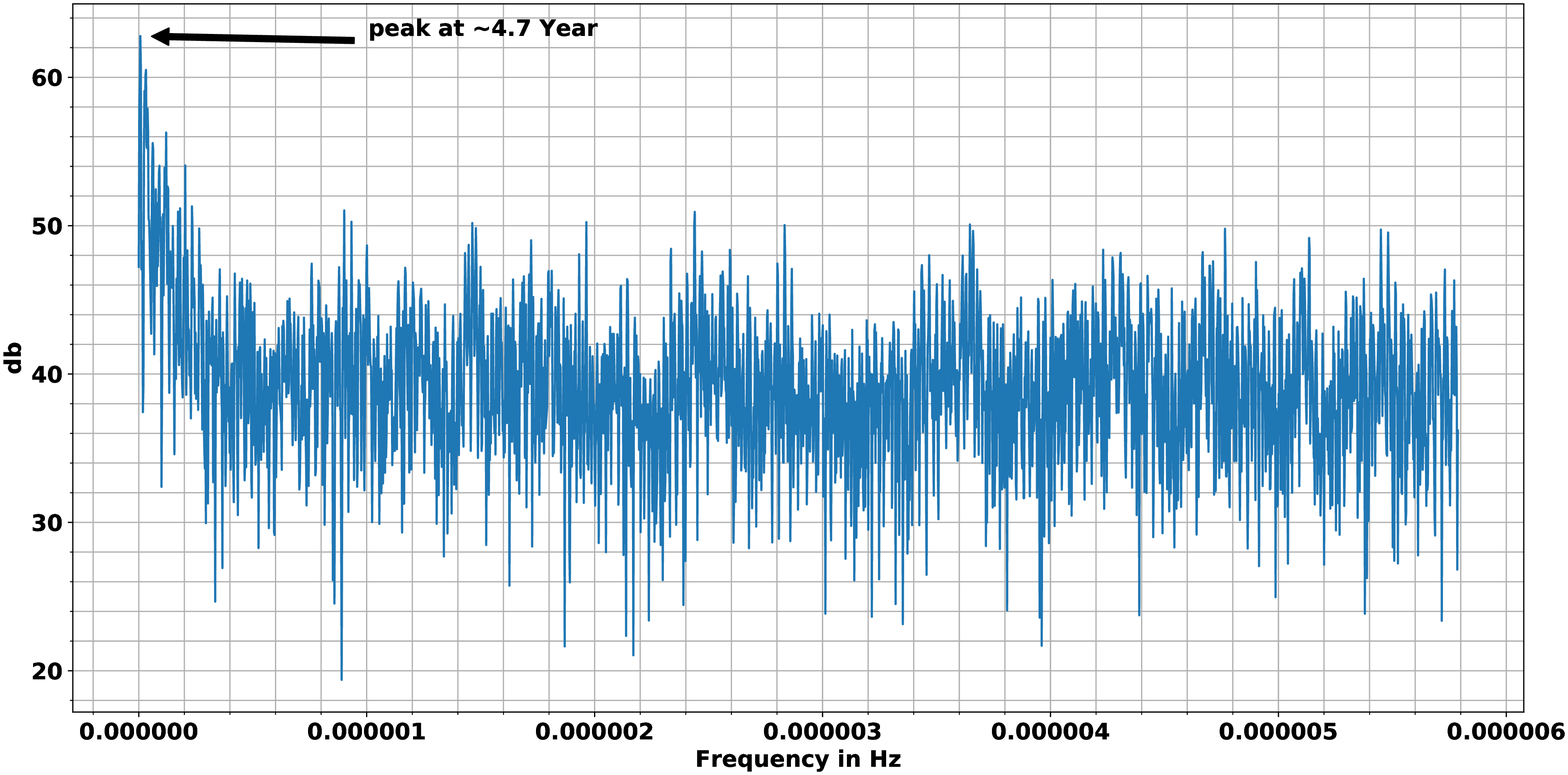}
\includegraphics[width=1\textwidth]{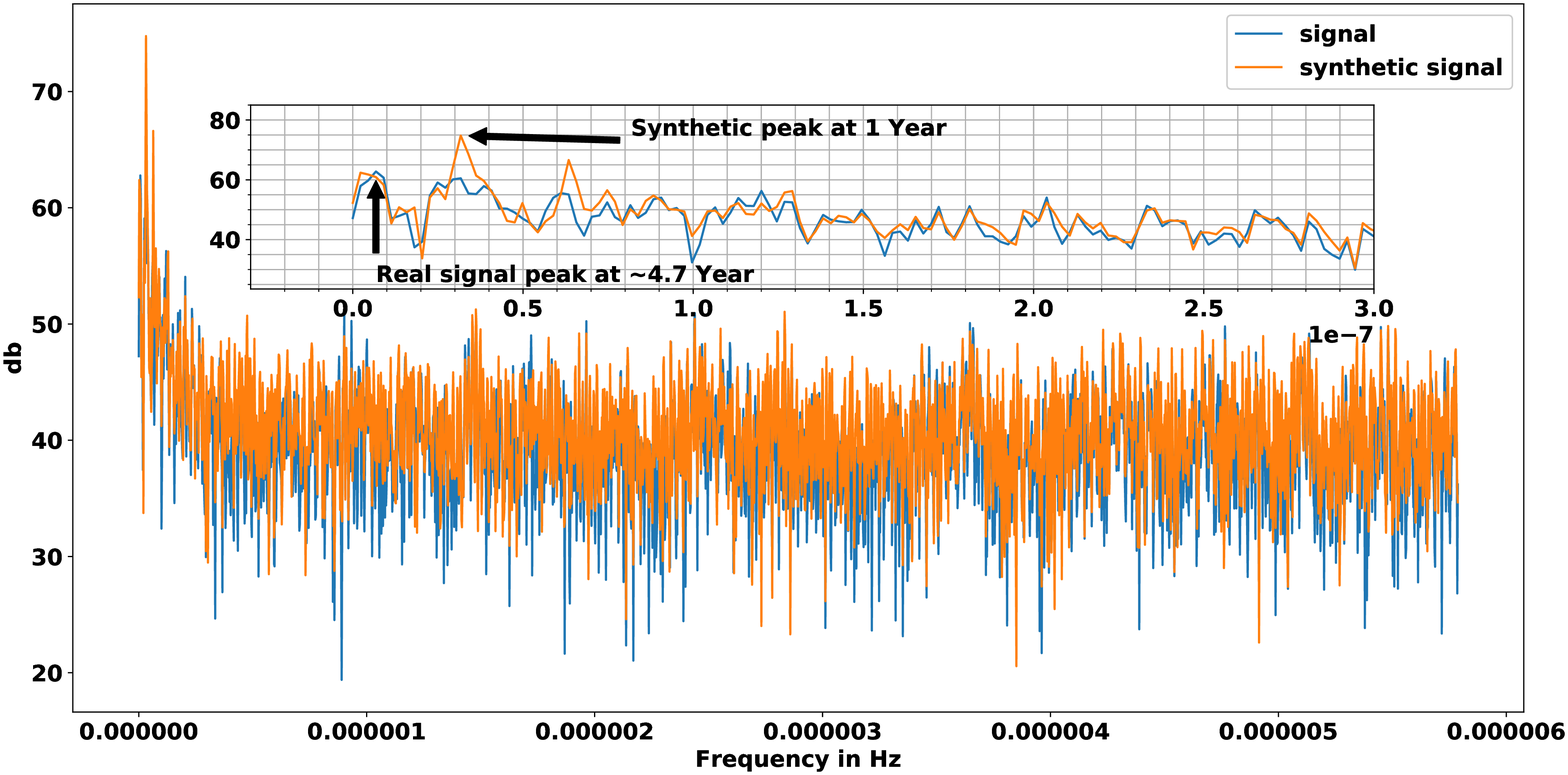}
\caption{Upper panel: the Power Spectrum Density (PSD). The most distinct peak $62.7$ db found at frequency $(6.7 \pm 0.23) \times10^{-9}$~Hz (corresponding to $\sim 4.7$ years). \\ 
Lower panel: our simulated signal (along side the real signal) with peak $74.7$ db at $(3.1 \pm 0.23) \times 10^{-8}$~Hz ($\simeq 1$ year).
}
\label{fig:welch}
\end{figure*}

To get the spectral density estimation we use Welch's method.
The result is presented in Fig. \ref{fig:welch}, as the Power Spectrum Density (PSD).
Here we can see the real signal peak at $(6.7 \pm 0.23) \times10^{-9}$~Hz and the synthetic 1~year signal as a peak at $(3.1 \pm 0.23) \times 10^{-8}$~Hz, as expected.
%

\section{Results}
\label{sec:resultss}

From the FFT analysis (Fig. \ref{fig:spec}), supported by the power spectrum (Fig. \ref{fig:welch}) we find a few peaks.
The most evident period in the signal is $P_1$ at $1735 \pm 115$~days with $27.5$~db. There is a secondary peak $P_2$ with a power of $26.1$ db and peaks $P_3$ to $P_5$ with a power of the order of $\simeq25$ db. We also notice that even though there is naturally high temporal resolution in the high frequencies (the resolution is $\propto \nu$, up to the frequency corresponding to the sampling of 1 day), there is a distinct gap between the strong peaks at the high frequencies region.
The peaks at the low frequencies are considerably stronger than the peaks at high frequencies.
From point $P_1$ onward to lower frequencies the signal declines.

The synthetic signal gives a peak of $37.5$~db at $362 \pm 5$~days, as expected.
The peak of the synthetic signal has its origin in a pure sinusoidal function with an amplitude of $\Delta V \simeq 0.1$ mag. Thus, its large power is a direct result of this large amplitude.
As expected, the power we obtain from our data is much weaker, since it relies on much smaller amplitudes.

Examining the PSD, we also find that the most distinct peak of $62.7$ db is at $\approx 4.7 \pm 0.3$ years.
As both methods provided the same periods, we conclude that the periodicities exist in the data.
Using Welsh method we obtain a flat signal of which most is contained below 45db. This indicates that the noise in the signal is white, namely frequency independent. The strong peak we obtained is almost 20db stronger than the level of white noise.
To show the significance of the peak quantitatively we go back to the FFT of the unified signal.
We calculated the statistical properties of the FFT intensity and found the mean intensity (the absolute value of the real and complex parts of the FFT) to be 3.8 and the standard deviation to be 4.1 (note that the zero-padding has no significance on the statistical properties of the properties but only on the frequency resolution).
The obtained strongest peak gives a normalized intensity of 23.7. Even the fifth peak has an intensity of 18.8.
We therefore conclude that the peaks are detected with high certainty.

\section{SUMMARY and Discussion}
\label{sec:summary}

Explaining the eruption of P~Cyg by mass transfer further supports
the conjecture that all major LBV eruptions are triggered by interaction
of an unstable LBV with a stellar companion.

The model of \cite{Kashi2010} predicted an orbital period of about 7 years while the longest period we found here was 4.7 years.
This gives rise to a few questions.

(a) \textit{Does the peak at 4.7 years indicate the existence of a binary companion?} There is no conclusive answer, but the chances are quite good. At first glance it may seem that in order for the periodiciy in the signal to be related to the light from a binary star the P~Cyg system needs to be almost edge-on, so that the companion will be obscured for most of the orbit. But this is not the case. The radius of the LBV is large, and we can add to it a wide region of dense wind that is optically thick in the visible range up to a considerable distance. It is therefore quite likely that a companion will shine for part of the orbit when it is on the observer side, and be obscured for the remainder of the orbit. 

(b) \textit{If not a binary, what else can the period indicate?} There are a few other possibilities: internal variation of the star, magnetic periodicity, unknown effects related to the LBV recovery from its eruption,
instabilities in the wind, and more.

(c) \textit{Is an orbit of 4.7 years compatible with the predictions of \cite{Kashi2010}?} 
It is possible, but not probable, as we now explain.
The prediction of \cite{Kashi2010} that a companion exists with an orbital period of $\simeq 7$ years comes from the assumption that mass accretion and mass loss from the LBV ended after its series of eruptions during 1654--1685. The period between the last two peaks was $\simeq 7$ years and no eruption has been observed since then.
A period of $4.7$ years is incompatible with the model of \cite{Kashi2010}, unless the orbit was reduced from $\simeq7$ years in 1685 to 4.7 years at the twentieth century.
In order to examine whether an orbit of 4.7 years is compatible with the predictions of \cite{Kashi2010} we repeat their calculations and find that in order to have an orbital period reduced to 4.7 years at the end of the twentieth century, the LBV should have transferred $\simeq0.28 \rmModot$ to the secondary. Since not all the mass lost from the LBV is accreted, the mass loss from the LBV over $\approx 350$ years would have to be $\approx 0.5 \rmModot$, or on average the mass loss rate would have to be $\approx 1.4 \times 10^{-3} \msyr$. While this is theoretically possible for an LBV, the accreted mass would have to emit its gravitational energy, at least partially as radiation that would have an observational signature.
However, no increase in luminosity as a result of these hypothetical eruptions has been recorded in the literature. Theoretically, every one or a few orbital periods an eruption might have occurred, but they all were weak and obscured by the ejecta from the strong eruptions observed in the seventeenth century.
Other mechanisms such as tidal interactions are weak and cannot account for the shortening of the period (see discussion in \citealt{Kashi2010}).
In summary, it is theoretically possible that we detected a companion star as predicted by \cite{Kashi2010}, but it requires the rather strong assumption of obscuration of eruptions that succeeded the ones observed.

Even if the 4.7 years period does not represent the companion proposed by \cite{Kashi2010},
their model still valid for three reasons.
First, it still explains almost perfectly the series of eruptions of P~Cyg in the seventeenth century.
Second, our present search for periodicity had a very small chance to find such a long period of $\approx 7$ years, since it only spanned 66 years. It is very clear from Fig. \ref{fig:spec} that a period of $7$ years is at the edge of the figure where the frequencies spread very thinly, so the observational duration is too short to allow finding such a long period.
Third, as discussed in section \ref{sec:observations}, the precision of the AAVSO observations could only detect the companion suggested by \cite{Kashi2010} for the most optimistic companion parameters and best observations precision.
Not detecting the companion suggested by \cite{Kashi2010} with the presently available data is therefore not a big surprise.

We also note that the ratio between the periods $\approx 4.7$ and $\simeq 7$ years is $2:3$, which might indicate some resonance with the suggested companion suggested by \cite{Kashi2010}.

We hope that AAVSO observers and other campaigns will continue to document the photometric variation of P~Cyg with an increasing precision, such that this exercise can be repeated in a few decades with a longer duration of observations.

\vspace{0.5cm}
We thank an anonymous referee for helpful comments that helped improving the paper.
We thank Noam Soker for very helpful discussions.
We acknowledge with thanks the variable star observations from the AAVSO International Database contributed by observers worldwide and used in this research.
AK acknowledges support from the R\&D authority in Ariel University and the Rector of Ariel University.
NK acknowledges Shota Rustaveli National Science Foundation (SRNSF grant No 218070).

\end{document}